\documentclass[lettersize,journal]{IEEEtran}
\usepackage{amsmath,amsfonts}
\usepackage{algorithmic}
\usepackage{algorithm}
\usepackage{array}
\usepackage[caption=false,font=normalsize,labelfont=sf,textfont=sf]{subfig}
\usepackage{textcomp}
\usepackage{stfloats}
\usepackage{url}
\usepackage{verbatim}
\usepackage{booktabs}
\usepackage{makecell}
\usepackage{multirow}
\usepackage{graphicx}
\usepackage{cite}

\newcounter{MYtempeqncnt}
\begin{document}

\title{A Practical Multi-Protocol Collaborative QKD Networking Scheme}

\author{Jia-Meng Yao,Qiong Li,Hao-Kun Mao,Ahmed A. Abd El-Latif
\thanks{Jiameng Yao, Qiong Li (corresponding author: qiongli@hit.edu.cn) and Hao-Kun Mao are with the Department
of the School of Computer Science and Technology, Harbin Institute of Technology, Harbin, China.Nan Chen is with the School of Foreign Languages, Harbin Institute of Technology, Harbin, China. Ahmed A. Abd El-Latif is with the EIAS Data Science Lab, College of Computer and Information Sciences, Prince Sultan University, Riyadh, Saudi Arabia, and with the Department of Mathematics \& Computer Science, Faculty of Science, Menoufia University, Egypt.}

\thanks{Manuscript received April 19, 2021; revised August 16, 2021.}}

\markboth{Journal of \LaTeX\ Class Files,~Vol.~14, No.~8, August~2021}%
{Shell \MakeLowercase{\textit{et al.}}: A Sample Article Using IEEEtran.cls for IEEE Journals}


\maketitle

\begin{abstract}
With the advancement of quantum computing technology, the security of classical public key cryptography is under serious threat. To guarantee security in the quantum era, Quantum Key Distribution (QKD) network technique has become a competitive solution and attracted wide attention. QKD networks can be classified into two main categories: measurement-device-dependent protocol-based network and measurement-device-independent protocol-based network. In measurement-device-dependent protocol-based networks, the transmitted information is available for all trusted relays. This means that all trusted relays are strongly trusted relays that require strict security control, which is very difficult to realize in real life. To address this issue, measurement-device-independent protocol-based networks reduce the proportion of strongly trusted relay nodes by introducing untrusted relays to alleviate the security control pressure. However, due to the higher key rate of measurement-device-dependent protocols over short distances, the communication capability of measurement-device-independent protocol-based networks has a large degradation compared to measurement-device-dependent protocol-based networks. Therefore, how to reduce the dependence of QKD networks on strong trusted relays without significantly affecting the communication capability of the network has become a major issue in the practicalization process of QKD networks. To address this issue, a novel Multi-Protocol Collaborative (MPC) networking cell is proposed in this paper. The QKD network built by the MPC networking cell reduces the dependence on strongly trusted relays by combining the two protocols to introduce weak trusted relays while maintaining the high communication capacity. What’s more, to further enhance the overall performance of the QKD network, an optimal topology design method is presented via the proposed flow-based mathematical model and optimization method. The simulation results show that the proposed scheme improves the communication capability by 37 times compared to measurement-device-independent schemes and reduces the strongly trusted relays by 23\% compared to measurement-device-dependent schemes. By addressing the challenge of reducing the dependence on strongly trusted relays without a significant reduction in communication capability, our work holds great significance in promoting the practicalization of QKD networks.
\end{abstract}


\section{Introduction}
\IEEEPARstart{W}{ith} the rapid evolution of information technology, an increasing amount of critical data, such as medical, financial, and government information, is being transmitted through networks. Therefore, it is very important to ensure the security of the information being communicated. Presently, prevailing encryption methods heavily rely on public key encryption algorithms that are based on computational security. However, the development of quantum computing technology has seriously threatened the security of classical public key cryptography~\cite{shor1999polynomial,grover1997quantum}.

Quantum Key Distribution (QKD) is a symmetric key distribution technology based on the fundamental principles of quantum mechanics. Heisenberg’s uncertainty principle and no-cloning theorem in quantum mechanics guarantee the security of QKD~\cite{flugge1999practical,heisenberg1927anschaulichen,wootters1982single}. QKD combined with the One-Time-Pad (OTP) encryption algorithm can achieve information-theoretic secure communication~\cite{shannon1949communication}. Therefore, the QKD technology holds great significance in the quantum era.

\subsection{Introduction to QKD Networks}
Since QKD can only provide point-to-point secure key distribution services, research on QKD networks is crucial to expanding the application scale of QKD technology and overcoming the limitations of communication distance. In recent years, numerous scholars have conducted relevant research on QKD networks, and the scale of experimental QKD networks has gradually expanded~\cite{chen2021integrated,chen2021twin,liu2021field}. The number of QKD network nodes has increased from 6 nodes to 56 nodes~\cite{razavi2018introduction,chen2020sending,liao2018satellite}, and the total length of the Beijing-Shanghai QKD network has reached over 2000 km~\cite{zhang2019quantum}. Futhermore, the launch of the Micius quantum satellite has further propelled the development of QKD networks, enabling quantum key generation across a distance of 7200km~\cite{liao2018satellite}. These experimental QKD networks have not only verified the feasibility of QKD networks but have also played a significant role in promoting the development of QKD networks. As a result, QKD networks are gradually transitioning from the experimental stage to the practical stage.

\subsection{Challenges and Our Motivation}
The existing QKD networks mainly consist of two types: measurement-device-dependent protocol-based QKD networks and measurement-device-independent protocol-based QKD networks. In measurement-device-dependent protocol-based networks, the transmitted information is available for all trusted relays. This means that all trusted relays are strongly trusted relays that require strict security control, which is very difficult to satisfy in practical applications. To address this issue, measurement-device-independent protocol-based networks reduce the proportion of strongly trusted relay nodes by introducing untrusted relays to alleviate the security control pressure. However, since the key generation rate of measurement-device-independent protocol is generally lower than that of measurement-device-dependent protocols in short distances, it results in limited communication capabilities that may be difficult to meet the demands of users. In conclusion, the aforementioned drawbacks of single protocols severely obstruct the practicalization of the QKD networks.

To address the above issues, scholars have conducted research on networking based on hybrid protocols, aiming to combine the advantages of different protocols. For example, in the domain of multi-protocol compatibility, the literature~\cite{cao2022software} focuses on the support of multiple QKD protocols using Software-Defined Networking (SDN) as the foundation. Additionally, ~\cite{fan2021measurement} proposed a nonstandalone (NSA) QKD device that can support multiple protocols (BB84 and MDI). These investigations on multi-protocol compatibility have demonstrated the feasibility of multi-protocol QKD networks. Furthermore, other scholars have conducted research on multi-protocol from a network design perspective. For example, in literature~\cite{cao2020mixed}, four schemes were proposed for arranging untrusted relays in hybrid QKD networks based on the link-splitting approach. However, it is important to note that the applicability of the link-splitting solution may be constrained by geographical considerations. Moreover,~\cite{wang2020topological} focuses on the issue of constructing multi-protocol QKD networks on classical networks. Although the above studies have achieved multi-protocol in QKD networks, the protocols still work independently. They generate keys separately and use them in isolation. In essence, the above schemes do not fundamentally achieve the goal of multi-protocol complementary collaboration.

\begin{figure*}[h]%
	\centering
	\includegraphics[width=1\textwidth]{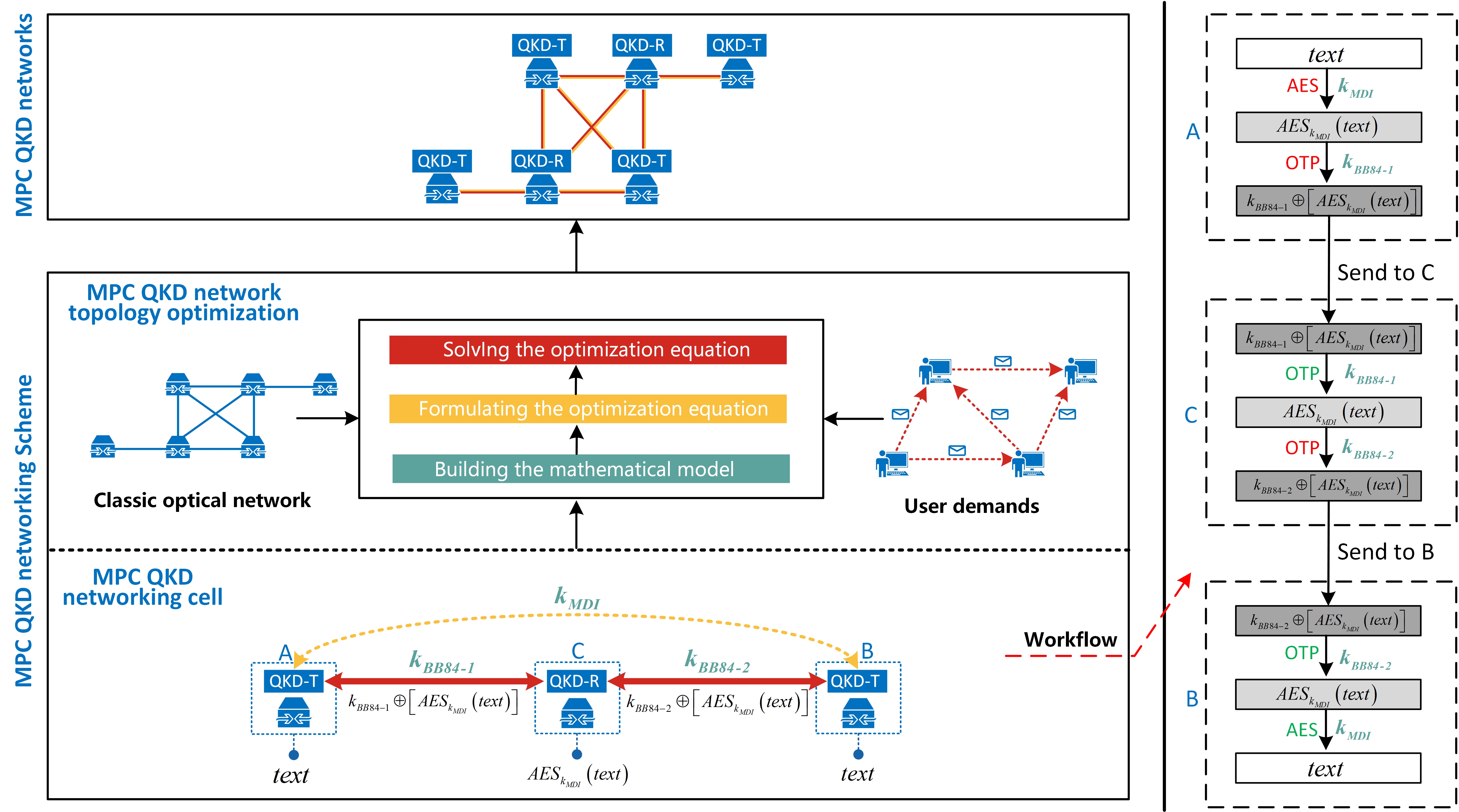}
	\caption{Framework of the proposed Multi-Protocol Collaborative (MPC) QKD networking scheme.}\label{fig1}
\end{figure*}

To address the above issues, this article proposes a multi-protocol collaborative networking scheme, as depicted in Figure~\ref{fig1}. First, a multi-protocol collaborative networking cell (MPC QKD networking cell) based on the NSA QKD device is proposed. In this paper the BB84 protocol is used for the measurement device dependent protocol and the MDI protocol is used for the measurement device independent protocol. Note that this is only used as an example in this paper, and other QKD protocols of the corresponding category can also be used. This networking cell collaboratively utilizes the keys from MDI and BB84 protocols through symmetric encryption algorithms, thereby leveraging the advantages of MDI’s weak dependence on strongly trusted relays and BB84’s high key generation rate. By doing so, the QKD network built by the MPC networking cell reduces the dependence on strongly trusted relays while maintaining high communication capacity. It is very significant to promote the practicalization of QKD networks. Moreover, to further enhance the performance of QKD networks, the issue of topology optimization that how to strategically deploy the MPC QKD networking cells on existing optical networks is also studied in this paper. Finally, the proposed MPC QKD networking scheme is compared with existing QKD networking schemes through simulation experiments. The results show that the proposed networking scheme reduces the dependence on strongly trusted relays without a significant reduction in communication capability, making it a more practical solution.

The main contributions of this paper are as follows:
\begin{itemize}
	\item{A practical multi-protocol collaboration (MPC) QKD networking cell is proposed.}
 
	\item{For the topology optimization problem of the proposed MPC QKD networks, the corresponding mathematical model and the topology optimization equation based on the optimization method are given.}

	\item{Simulation experiments were conducted to compare the performance of the proposed MPC NSA QKD networks with other existing QKD networks.}
\end{itemize}

The remainder of this paper is organized as follows. The MPC QKD networking cell proposed in this paper is introduced in detail in Section 2. The mathematical model and the topology optimization equations of the MPC QKD networks are given in Section 3. Simulation results and detailed analysis are provided in Section 4. Finally, the summary is drawn in Section 5.

\section{MPC QKD Networking Cell}
\subsection{Work flow}
The multi-protocol collaborative (MPC) networking cell proposed in this paper is designed based on the NSA QKD device proposed in~\cite{fan2021measurement}. The NSA QKD device can support both the BB84 and MDI protocols, which provides a solid physical device foundation for multi-protocol collaboration. The MPC QKD networking cell can comprehensively take advantage of the high key rate of BB84 and the weak dependence on strongly trusted relays of MDI.

The working principle of the MPC QKD networking cell is depicted in Figure~\ref{fig1}. An NSA QKD device is installed on the A-C-B structure. Nodes A and B can generate quantum secure keys based on the MDI protocol through C as an untrusted relay, and the generated keys is denoted as ${k_{MDI}}$. In addition, quantum secure keys can also be generated between nodes A and C or nodes B and C based on the BB84 protocol, and the generated keys are denoted as ${k_{BB84 - 1}}$ and ${k_{BB84 - 2}}$ respectively. In the proposed MPC QKD networking cell, when node A intends to transmit a message to node B, the workflow is as follows:

\begin{enumerate}
\item The source node A encrypts the message $text$ using a symmetric encryption algorithm based on the key $k_{MDI}$ generated between it and node B. In this paper, the AES symmetric encryption algorithm is used for this encryption. It should be noted that other symmetric encryption algorithms can also be utilized to perform this encryption. The ciphertext after this encryption step is denoted as ${AES_{{k_{MDI}}}}\left( {text} \right)$.
\item The source node A further encrypts the ciphertext ${AES_{{k_{MDI}}}}\left( {text} \right)$ obtained in step 1 using the OTP algorithm based on the key ${k_{BB84 - 1}}$ generated between it and node C. The ciphertext after this encryption step is denoted as ${k_{BB84 - 1}} \oplus \left[ {AE{S_{{k_{MDI}}}}\left( {text} \right)} \right]$.
\item The source node A transmits the ciphertext ${k_{BB84 - 1}} \oplus \left[ {AE{S_{{k_{MDI}}}}\left( {text} \right)} \right]$ to node C through AC channel.
\item Upon receiving the ciphertext ${k_{BB84 - 1}} \oplus \left[ {AE{S_{{k_{MDI}}}}\left( {text} \right)} \right]$, node C firstly decrypts it using the key ${k_{BB84 - 1}}$ generated between it and node A, obtaining the ${AES_{{k_{MDI}}}}\left( {text} \right)$. Subsequently, ${AES_{{k_{MDI}}}}\left( {text} \right)$ is re-encrypted using OTP algorithm based on the key ${k_{BB84 - 2}}$ generated between node C and node B, obtaining the ciphertext ${k_{BB84 - 2}} \oplus \left[ {AE{S_{{k_{MDI}}}}\left( {text} \right)} \right]$.
\item The node C transmits the ciphertext ${k_{BB84 - 2}} \oplus \left[ {AE{S_{{k_{MDI}}}}\left( {text} \right)} \right]$ to node B through CB channel.
\item Upon receiving the ciphertext ${k_{BB84 - 2}} \oplus \left[ {AE{S_{{k_{MDI}}}}\left( {text} \right)} \right]$, node B decrypts it using the key ${k_{BB84 - 2}}$ generated between it and node C, obtaining the ${AES_{{k_{MDI}}}}\left( {text} \right)$.
\item Node B then decrypts ${AES_{{k_{MDI}}}}\left( {text} \right)$ using the key $k_{MDI}$ generated between it and node A, obtaining the plaintext message $text$ sent from node A.
\end{enumerate}

As evident from the above workflow, the information transferred at the relay node is ${AES_{{k_{MDI}}}}\left( {text} \right)$ which is encrypted using the symmetric encryption algorithm based on the key $k_{MDI}$. Given the high security of both $k_{MDI}$ and symmetric encryption algorithms, the security of the information at the relay node is significantly enhanced, which we call a weak trusted relay. Compared with the strong trusted relay, the weak trusted relay significantly reduces the difficulty of trusted supervision in practical applications and has stronger practicality. A detailed security analysis will be provided in the subsequent section. In addition, since the key $k_{MDI}$ primarily serves as an auxiliary key, the transmission rate mainly depends on the key generation rate of BB84. As a result, the MPC QKD networking cell reduces the network's dependence on strongly trusted relays without significantly affecting communication capability by introducing such weakly trusted relays, which is of great significance in promoting the practicality of QKD networks.

\subsection{Security Analysis}
This section provides the security analysis of the MPC QKD networking cell. As is known from Section 2.1 that the information exposed at channels AC, CB and relay node C is ${k_{BB84 - 1}} \oplus \left[ {AE{S_{{k_{MDI}}}}\left( {text} \right)} \right]$, ${k_{BB84 - 2}} \oplus \left[ {AE{S_{{k_{MDI}}}}\left( {text} \right)} \right]$, and ${AES_{{k_{MDI}}}}\left( {text} \right)$, respectively. Firstly, the message ${k_{BB84 - 1}} \oplus \left[ {AE{S_{{k_{MDI}}}}\left( {text} \right)} \right]$ transmitted through channel AC is obtained by encrypting the ciphertext ${AES_{{k_{MDI}}}}\left( {text} \right)$ using ${k_{BB84 - 1}}$ with the OTP encryption algorithm. Therefore, according to~\cite{shannon1949communication} the message $text$ possesses information-theoretic security on the channel AC. Similarly, message $text$ also possesses information-theoretic security on channel CB. The information exposed at the relay node C is ${AES_{{k_{MDI}}}}\left( {text} \right)$, so its security mainly depends on security of the AES symmetric encryption algorithm and the key ${k_{MDI}}$. Since the security of the key ${k_{MDI}}$, generated by QKD, is based on the principle of quantum mechanics, it is considered to be unconditionally secure. Furthermore, according to the analysis of~\cite{bonnetain2019quantum}, the AES-256 symmetric encryption algorithm has post-quantum security. As a result, the information at relay node C achieves post-quantum security, representing a significant improvement in security compared to strongly trusted relays, which means it substantially reduces the dependence on strongly trusted relays.

\subsection{Advantage analysis}
In this section, the analysis is conducted to examine the advantages of the MPC networking cell based QKD networks compared to the BB84 and MDI networking cell based QKD networks in terms of communication capability and dependence on strongly trusted relays.

\begin{figure}[h]%
	\centering
	\includegraphics[width=0.45\textwidth]{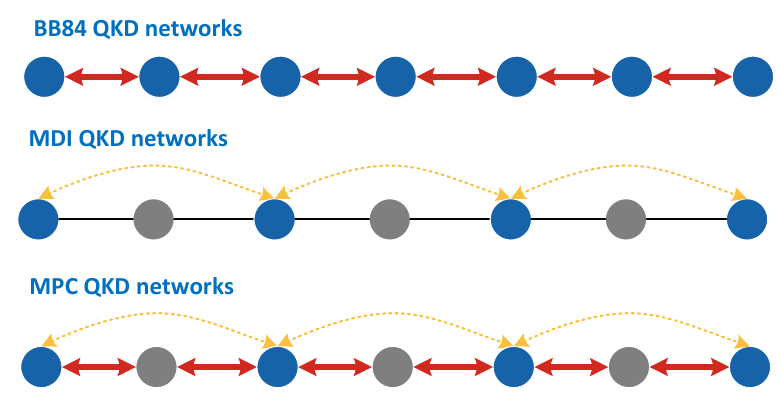}
	\caption{Comparision of linear BB84 QKD networks, MDI QKD networks, and MPC QKD networks.}\label{fig2}
\end{figure}

{\textbf{BB84 QKD networks: }}  In QKD networks based on the BB84 networking cell, the security of information transmitted at the relay nodes is seriously threatened, resulting in a strong dependence on strongly trusted relays. In contrast, the MPC QKD networking cell based QKD networks transmit information at the relay nodes in the form of symmetric encryption, which is weak trusted relays. Consequently, the MPC QKD networks have a weaker dependence on strongly trusted relays. For example, in the linear network with $n$ nodes shown in Figure~\ref{fig2} (more complex network cases will be analyzed in Section 4), the number of strongly trusted relays required for the BB84 QKD networks is $n-2$, while for the MPC QKD networks is $\left\lfloor {\frac{{n - 2}}{2}} \right\rfloor $. This indicates that the proposed networking cell significantly reduces the dependence on strongly trusted relays during information transmission. In terms of the transmission rate, the proposed networking cell requires additional time for MDI key generation, resulting in a slightly lower transmission rate compared to BB84. The time ratio allocated for generating MDI keys can be adjusted according to specific requirements. Assuming that $1/10$ of the time is used for the MDI key generation, the transmission rate of the MPC networking cell would be $9/10$ of that of BB84. From the above analysis, it can be concluded that the proposed networking cell based QKD networks achieve a significant improvement in reducing the dependence on strongly trusted relays compared to the BB84 QKD networks, at the cost of a slight reduction in the communication rate. 

{\textbf{MDI QKD networks: }} In terms of the dependence on strongly trusted relays, both the MPC QKD networks and the MDI QKD networks have a weak dependence on strongly trusted relays according to the security analysis in the previous section. In terms of transmission rates, the MPC networks outperform the MDI networks significantly, as the transmission rate of the MPC networking cell depends on the key rate of BB84 protocol.

In summary, the MPC QKD networks not only have a weak dependence on strongly trusted relays but also has a high communication rate. It effectively addresses the issues of strong dependence on strongly trusted relays of QKD networks without significant reduction in communication capacity. Therefore, the proposed MPC QKD networking cell holds significant practical application value in QKD network.

\section{Topology optimization of MPC QKD networks}
To further enhance the performance of QKD networks, this section focuses on the problem of topology optimization (optimal deployment of the networking cell on existing optical networks) for the MPC QKD networks. The mathematical model and topology optimization equations of the MPC QKD networks are presented in Section 3.1 and Section 3.2, respectively.

\subsection{Mathematical model}
To address the topology optimization problem, this section presents the mathematical model of MPC QKD networks by abstracting the key elements in the network, such as nodes, links, communication demands, and key bandwidth. The mathematical model is mainly based on the flow model proposed in our previous work~\cite{li2020mathematical}. However, since the MPC QKD networks are based on the proposed MPC QKD networking cell, certain modifications need to be made to the mathematical model proposed in~\cite{li2020mathematical}.

First, we abstract the network topology into a graph $G = \left( {V,E} \right)$, where $V$ represents the set of nodes and $E$ represents the set of links in the network. In addition, since the MPC QKD networks are based on NSA QKD devices, the deployment of an NSA QKD device requires two adjacent edges. Therefore, we use the set $\widehat E$ to represent all the structures in this network that can be used to deploy the NSA QKD devices. For simplicity, we introduce the description in~\cite{wang2020topological}: the edges in sets $E$ and $\widehat E$ are referred to as client-to-client (C2C) and client-server-client (CSC) edges, respectively. Moreover, the flows on C2C edges and CSC edges are referred to as C2C-flows and CSC-flows, respectively. The notations and definitions used in the MPC QKD networks are described in Table~\ref{tab1}.

\begin{table*}[h]
	\begin{center}
		
			\caption{Notations and Definitions}\label{tab1}%
			
			\begin{tabular}{@{}lll@{}}
				\toprule
			      Notation  & Explanation & Value \\
				\midrule
				$\left\{ {{D^{s,t}}\left| {s,t \in V} \right.} \right\}$   & Average communication demands of communication pair $\left( {s,t} \right)$  & $R_0^ + $\\
				
				$T\left( s \right)$   & Credibility of node $s$  & $\left\{ {0,1} \right\}$\\
				
				$q$   & Trusted control cost  & $\left[ {0, + \infty } \right]$\\
				
				${S_{\left( {u,v} \right)}}$   & Number of equivalent QKD devices on edge $\left( {u,v} \right)$  & $N$\\
				
				${\hat S_{\left( {u,p,v} \right)}}$   & Number of actual QKD devices on edge $\left( {u,p,v} \right)$  & $N$\\

                    ${R_{B\left( {u,v} \right)}}$ & BB84-based key generation rate on edge $\left( {u,v} \right)$ & $R_0^ + $ \\
                    
                    ${R_{M\left( {u,p,v} \right)}}$ & MDI-based key generation rate on edge $\left( {u,p,v} \right)$ & $\left[ {0, + \infty } \right]$ \\
                    
                    ${R_{\left( {u,p,v} \right)}}$ & Key bandwidth of MPC QKD networking cell on edge $\left( {u,p,v} \right)$ & $\left[ {0, + \infty } \right]$ \\
                    
                   $\beta$ & Working time ratio of BB84 and MDI protocols in MPC QKD networking cell & $\left[ {0,1} \right]$ \\
                   
                    $\tau$ & Working time percent of MPC QKD networking cell & $\left[ {0,1} \right]$ \\
                    
                    $f_{\left( {u,v} \right)}^{s,t}$ & C2C-flows on the edge $\left( {u,v} \right)$ during node pair $\left( {s,t} \right)$ communication &  $R_0^ + $ \\
                    
                    $\hat f_{\left( {u,p,v} \right)}^{s,t}$ & CSC-flows on the edge $\left( {u,p,v} \right)$ with node $p$ acting as an untrusted relay during node pair $\left( {s,t} \right)$ communication & $R_0^ + $ \\
				\bottomrule
			\end{tabular}
	
		\end{center}
\end{table*}

Since the MPC QKD networks are based on the proposed MPC QKD networking cell, some of these definitions require special explanation. ${\widehat S_{\left( {u,p,v} \right)}}$ represents the number of actually deployed NSA QKD devices on edge $\left( {u,p,v} \right)$. As this CSC edge can communicate not only based on the MPC QKD networking cell but also based on the BB84 protocol, the edges $\left( {u,p} \right)$ and $\left( {p,v} \right)$ can also obtain the key bandwidth after deploying the QKD devices. To describe the total key bandwidth on such C2C edges, ${S_{\left( {u,v} \right)}}$ is introduced to represent the number of equivalent QKD devices on the C2C edge. If an actual QKD device is deployed on the edge $\left( {u,p,v} \right)$, then the number of equivalent QKD devices on the edge $\left( {u,p} \right)$ and $\left( {p,v} \right)$ will be increased by 1. Therefore, the number of equivalent QKD devices on each C2C edge can be calculated by \eqref{eq1}

\begin{equation}
	\label{eq1}
	{S_{\left( {u,v} \right)}} = \sum\limits_{t \in V} {{{\hat S}_{(u,v,t)}}}  + \sum\limits_{s \in V} {{{\hat S}_{(s,u,v)}}} 
\end{equation}

In addition, since the proposed MPC NSA QKD networks is mainly based on the MDI and BB84 protocol, ${R_{B\left( {u,v} \right)}}$ and ${R_{M\left( {u,p,v} \right)}}$ is used to denotes the BB84-based key generation rate of the edge $\left( {u,v} \right)$ and the MDI-based key generation rate of the edge $\left( {u,p,v} \right)$ respectively, which can be calculated by the GLLP according to the link length~\cite{gottesman2004security}. Due to the MPC QKD networks requires two protocols to work collaboratively, the working time allocation ratio of BB84 and MDI protocols is denoted as $\beta$. In addition, since QKD devices can work based on both the proposed MPC QKD networking cell (where the relay nodes are weakly trusted) and the BB84 protocol (where the relay nodes are fully trusted), $\tau$ is used to denote the percent of working time of the MPC networking cell. Therefore, the key bandwidth on the CSC edge $\left( {u,p,v} \right)$ when the QKD devices work based on the MPC QKD networking cell is denoted by ${R_{\left( {u,p,v} \right)}}$. And the calculation formula for ${R_{\left( {u,p,v} \right)}}$ is as follows \eqref{eq2}:

\begin{equation}
	\label{eq2}
	 {R_{\left( {u,p,v} \right)}} = \beta \left[ {{R_{B\left( {u,p} \right)}}{R_{B\left( {p,v} \right)}}/\left( {{R_{B\left( {u,p} \right)}} + {R_{B\left( {p,v} \right)}}} \right)} \right]
\end{equation}

We use $f_{\left( {u,v} \right)}^{s,t}$ to represent the flows over the edge $\left( {u,v} \right)$ when the node pair $\left( {s,t} \right)$ communicates, noting that the edge provides the keys directly based on the BB84 protocol at this time. $\hat f_{\left( {u,p,v} \right)}^{s,t}$ is used to represent the flows over the edge $\left( {u,p,v} \right)$ when the node pair $\left( {s,t} \right)$ communicates, noting that the edge works based on the MPC networking cell at this time.

\subsection{Topology optimization}
Based on the mathematical model presented in Section 3.1, this section focuses on the problem of topology optimization for the MPC QKD networks. The topology optimization process mainly includes the following objective functions and constraints. 

\begin{itemize}
\item Objective function
\end{itemize}

Optimizing the performance of communication is the main objective when designing the topology of QKD networks. Considering the varying communication demands of each communication pair in the network, our aim is to maximize the satisfaction of these demands. Therefore, the concept of satisfaction degree of communication demand (SoD) proposed in~\cite{wang2020topological} is introduced. The definition of SoD is as follows:
{\textbf{SoD: }} Suppose the communication demand of node pair $\left( {s,t} \right)$ is ${D^{s,t}}$ and the traffic from source node $s$ to destination node $t$ during the communication is  ${A^{s,t}}$. Then the satisfaction degree of communication demand of node pair $\left( {s,t} \right)$ can be denoted as ${B^{s,t}}$:

\begin{equation}
	\label{eq3}
	 {B^{s,t}} = \frac{{{A^{s,t}}}}{{{D^{s,t}}}}
\end{equation}

where ${A^{s,t}}$ can be calculated as the difference between the flow out of the source node $s$ and the flow into the source node $s$ as follows:

\begin{equation}
	\label{eq4}
	 {A^{s,t}} = \sum\limits_{v \in V} {\left[ {f_{\left( {s,v} \right)}^{s,t} - f_{\left( {v,s} \right)}^{s,t}} \right]}  + \sum\limits_{p,v \in V} {\left[ {\widehat f_{\left( {s,p,v} \right)}^{s,t} - \widehat f_{\left( {v,p,s} \right)}^{s,t}} \right]}
\end{equation}

When setting the objective function, it is hoped that the demands of each communication pairs could be satisfied. First, the SoD of all communication pairs is solved and get its minimum value. Then maximize it in the objective function. Therefore, the objective function equation is as follows:

\begin{equation}
	\label{eq5}
	 \mathop {\max }\limits_{F,\hat F,\widehat S,T} \mathop {\min }\limits_{s,t \in V} \frac{{{A^{s,t}}}}{{{D^{s,t}}}}
\end{equation}

\begin{itemize}
\item Bandwidth Constraint
\end{itemize}

Since communication in QKD networks heavily relies on keys, the flows on each edge should not exceed the key bandwidth of the edge. In the MPC QKD networks, the flows consist of  C2C-flows (based on the BB84 protocol) and CSC-flows(based on the MPC networking cell). Therefore, it essential for both the C2C-flows and the CSC-flows to satisfy the key bandwidth constraint. Thus, the key bandwidth constraint can be expressed as follows:

\begin{equation}
    \label{eq6}
    \begin{array}{l}
\forall \left( {u,v} \right) \in E,0 \le \sum\limits_{s,t \in V} {\left[ {f_{\left( {u,v} \right)}^{s,t} + f_{\left( {v,u} \right)}^{s,t}} \right]} \\
{\rm{                     }} \le {S_{\left( {u,v} \right)}} \cdot \frac{{\left( {1 - \tau } \right){R_{\left( {u,v} \right)}}}}{2} + {S_{\left( {v,u} \right)}} \cdot \frac{{\left( {1 - \tau } \right){R_{\left( {v,u} \right)}}}}{2}
\end{array}
\end{equation}

\begin{equation}
	\label{eq7}
	 \begin{array}{l}
\forall \left( {u,p,v} \right) \in \widehat E,0 \le \sum\limits_{s,t \in V} {\left[ {\widehat f_{\left( {u,p,v} \right)}^{s,t} + \widehat f_{\left( {v,p,u} \right)}^{s,t}} \right]} \\
{\rm{                         }} \le {\widehat S_{\left( {u,p,v} \right)}} \cdot \tau  \cdot {R_{\left( {u,p,v} \right)}} + {\widehat S_{\left( {v,p,u} \right)}} \cdot \tau  \cdot {R_{\left( {v,p,u} \right)}}
\end{array}
\end{equation}

where:

\begin{equation}
	\label{eq8}
	 {R_{\left( {u,p,v} \right)}} = \beta \left[ {{R_{B\left( {u,p} \right)}}{R_{B\left( {p,v} \right)}}/\left( {{R_{B\left( {u,p} \right)}} + {R_{B\left( {p,v} \right)}}} \right)} \right]
\end{equation}

\begin{equation}
	\label{eq8}
	 {S_{\left( {u,v} \right)}} = \sum\limits_{t \in V} {{{\hat S}_{(u,v,t)}}}  + \sum\limits_{s \in V} {{{\hat S}_{(s,u,v)}}} 
\end{equation}

\begin{itemize}
\item Flow Constraint
\end{itemize}

During the communication process, it is crucial for all nodes, besides the source and destination nodes, to ensure that the traffic flowing into them is equal to the traffic flowing out of them.

\begin{equation}
	\label{eq9}
	  \begin{array}{l}
\forall s,t \in V,\forall u \in \left( {V - \left\{ {s,t} \right\}} \right),\\
\sum\limits_{p,v \in V} {\left[ {\left( {f_{\left( {u,v} \right)}^{s,t} + \widehat f_{\left( {u,p,v} \right)}^{s,t}} \right) - \left( {f_{\left( {v,u} \right)}^{s,t} + \widehat f_{\left( {v,p,u} \right)}^{s,t}} \right)} \right] = 0} 
\end{array} 
\end{equation}

\begin{itemize}
\item Cost Constraint
\end{itemize}

Considering the expensive price of QKD devices and the cost of trusted nodes  control in the QKD networks, it is necessary to consider the cost constraints when designing the QKD network topology~\cite{wang2020topological}. The cost of QKD devices is determined by the total number of devices deployed in the QKD networks, while the cost of trusted control is determined by the number of trusted nodes. In this paper, it is considered that trusted control is required if a node is located at the endpoint of a CSC edge. Thus, the cost constraint can be expressed as follows:

\begin{equation}
	\label{eq10}
	  \sum\limits_{\left( {u,p,v} \right) \in \hat E} {{{\hat S}_{\left( {u,p,v} \right)}} + q \cdot \sum\limits_{s \in V} {T\left( s \right)}  - C \le 0}
\end{equation}

where $C$ is the total cost of network construction.

\begin{itemize}
\item Optimization equations
\end{itemize}

In summary, the topology optimization equation of the MPC QKD networks is as ~\eqref{eq11}.

\begin{figure*}[!t]
\normalsize
\setcounter{MYtempeqncnt}{\value{equation}}
\setcounter{equation}{11}
\begin{equation}
\label{eq11}
\begin{array}{l}
\mathop {\max }\limits_{F,\widehat F,\widehat S,T} \left( {\mathop {\min }\limits_{s,t \in V} \frac{{{A^{s,t}}}}{{{D^{s,t}}}}} \right)\\
s.t.\\
\forall \left( {u,v} \right) \in E,0 \le \sum\limits_{s,t \in V} {\left[ {f_{\left( {u,v} \right)}^{s,t} + f_{\left( {v,u} \right)}^{s,t}} \right] \le {S_{\left( {u,v} \right)}} \cdot \frac{{\left( {1 - \tau } \right){R_{\left( {u,v} \right)}}}}{2} + {S_{\left( {v,u} \right)}} \cdot \frac{{\left( {1 - \tau } \right){R_{\left( {v,u} \right)}}}}{2}} \\
\forall \left( {u,p,v} \right) \in \widehat E,0 \le \sum\limits_{s,t \in V} {\left[ {\widehat f_{\left( {u,p,v} \right)}^{s,t} + \widehat f_{\left( {v,p,u} \right)}^{s,t}} \right] \le {{\widehat S}_{\left( {u,p,v} \right)}} \cdot \tau  \cdot {R_{\left( {u,p,v} \right)}} + {{\widehat S}_{\left( {v,p,u} \right)}} \cdot \tau  \cdot {R_{\left( {v,p,u} \right)}}} \\
\forall \left( {u,p,v} \right) \in \widehat E,{R_{\left( {u,p,v} \right)}} = \beta \left[ {{R_{B\left( {u,p} \right)}}{R_{B\left( {p,v} \right)}}/\left( {{R_{B\left( {u,p} \right)}} + {R_{B\left( {p,v} \right)}}} \right)} \right]\\
\forall \left( {u,v} \right) \in E,{S_{\left( {u,v} \right)}} = \sum\limits_{t \in V} {{{\hat S}_{(u,v,t)}}}  + \sum\limits_{s \in V} {{{\hat S}_{(s,u,v)}}} \\
\forall s,t \in V,\forall u \in \left( {V - \left\{ {s,t} \right\}} \right),\sum\limits_{p,v \in V} {\left[ {\left( {f_{\left( {u,v} \right)}^{s,t} + \widehat f_{\left( {u,p,v} \right)}^{s,t}} \right) - \left( {f_{\left( {v,u} \right)}^{s,t} + \widehat f_{\left( {v,p,u} \right)}^{s,t}} \right)} \right] = 0} \\
\sum\limits_{\left( {u,p,v} \right) \in \hat E} {{{\hat S}_{\left( {u,p,v} \right)}} + q \cdot \sum\limits_{s \in V} {T\left( s \right)}  - C \le 0} \\
\forall s,t \in V,{A^{s,t}} = \sum\limits_{v \in V} {\left[ {f_{\left( {s,v} \right)}^{s,t} - f_{\left( {v,s} \right)}^{s,t}} \right]}  + \sum\limits_{p,v \in V} {\left[ {\widehat f_{\left( {s,p,v} \right)}^{s,t} - \widehat f_{\left( {v,p,s} \right)}^{s,t}} \right]} 
\end{array}
\end{equation}
\setcounter{equation}{\value{MYtempeqncnt}}
\hrulefill
\vspace*{4pt}
\end{figure*}

The decision variables of the above equation are $F$, $\widehat F$, $\widehat S$, $T$. The QKD networks can be constructed based on the optimal number of deployed QKD devices on each edge.

\section{Performance Evaluation}
\subsection{Experimental parameter settings}
\begin{figure}[h]%
	\centering
	\includegraphics[width=0.5\textwidth]{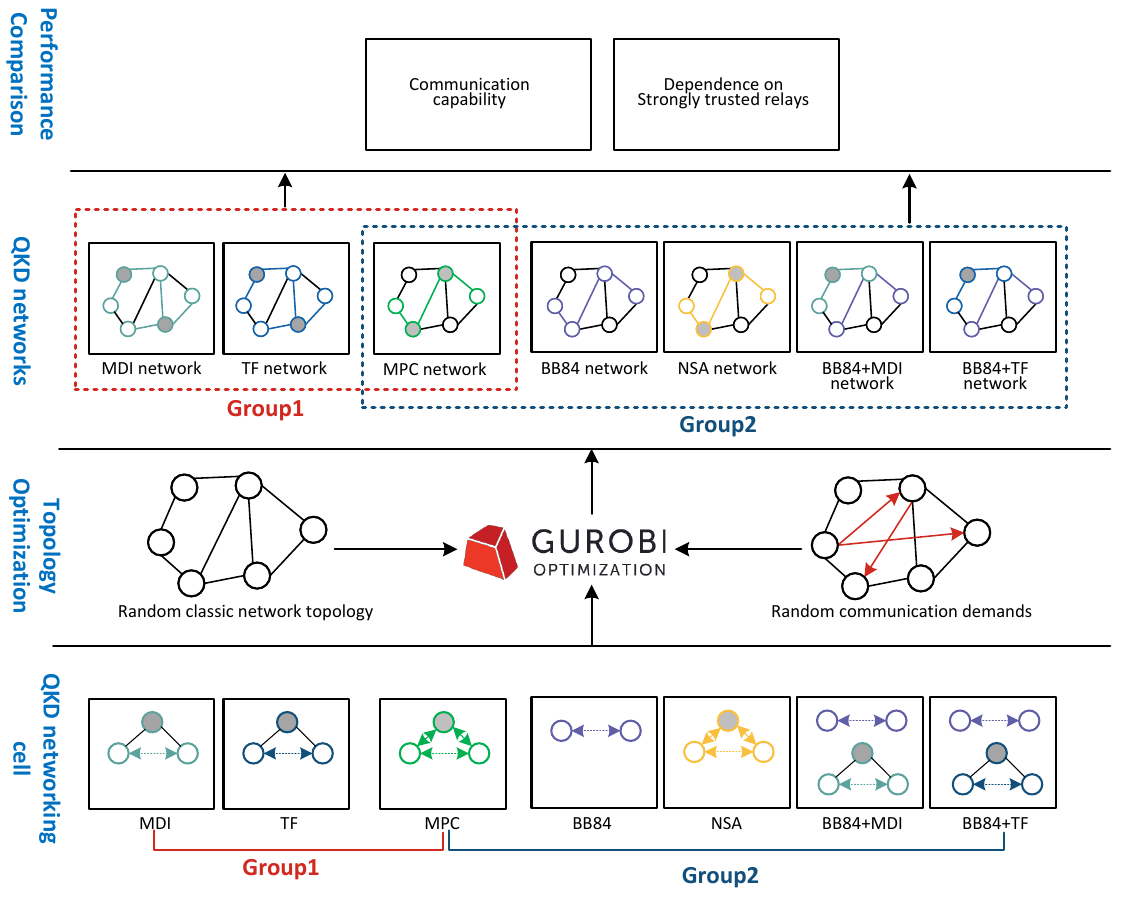}
	\caption{Workflow of the experiment}\label{fig3}
\end{figure}

In this section, the performance of the proposed MPC QKD networking scheme is analyzed through simulation experiments. The MPC QKD networks are compared with existing QKD networks in terms of communication capability and their dependence on strongly trusted relays. The workflow of the experiment is shown in Figure~\ref{fig3}. With the given classical network topology and communication demands, the QKD networks are constructed based on different networking cells by topology optimization, and then the performance of these QKD networks is compared.

The existing networking cells mainly include the following: MDI networking cell and Twin-field (TF) networking cell (based on untrusted relay protocol), BB84 networking cell (based on the trusted relay protocol), MDI-BB84 hybrid networking cell and TF-BB84 hybrid networking cell (based on multi-protocols). In addition, since the MPC QKD networking cell is based on the NSA QKD device, the performance of the NSA networking cell based on the original working mechanism in~\cite{fan2021measurement} is also compared in the experiments. The above networking cells are divided into two groups for comparison in this experiment. Group 1 is composed of the networking cells in which all the communication flows in the network are CSC-flows. Group 1 includes MDI networking cell, TF networking cell, and the proposed MPC networking cell when the value of $\tau$ is set to 1. Group 2 is composed of the networking cells in which C2C-flows are present in the communication. Group 2 includes the BB84 networking cell, the NSA networking cell, the BB84-MDI hybrid networking cell, the BB84-TF hybrid networking cell, and the proposed MPC networking cell ($\tau$ is set to 0.5 for a fair comparison with the NSA networking cell).

The main evaluation indicators of concern in this article are the communication capability and the dependence on strongly trusted relays of the QKD networks. For the communication capability, we use the satisfaction degree of communication demand (SoD) as the evaluation indicator (described in Section 3.2). In terms of the dependence on strongly trusted relays, the proportion of the CSC-flows, $CSC\_P = \sum {\widehat f} /\left( {\sum {\widehat f}  + \sum f } \right)$,is used as the evaluation indicator in this experiment.

To ensure fairness in the experiment, 15 networks with varying numbers of nodes \{10, 13, 15, 18, 20, 23, 25, 28, 30, 33, 35, 38, 40, 43, 45\} are simulated. Additionally, for each network scale, 10 groups of network topologies are randomly generated, and their average values are calculated for comparative analysis. In order to resemble realistic scenarios, the number of edges is set to 1.5 times the number of nodes, and the length of edges is set within the range of [10,250] km. For the communication demands, one-third of the nodes in the network are randomly selected to communicate with one-fifth of the nodes in the network in this experiment. The communication demands are set within the range of [100,300] kbps. The cost of trusted control $q$ is set to 100. And the total cost of network construction is set to 10000. The Gurobi solver is used to solve the topology optimization equations of the QKD networks.

\subsection{Simulation Results and Analysis}
\begin{figure*}[h]%
	\centering
	\includegraphics[width=1\textwidth]{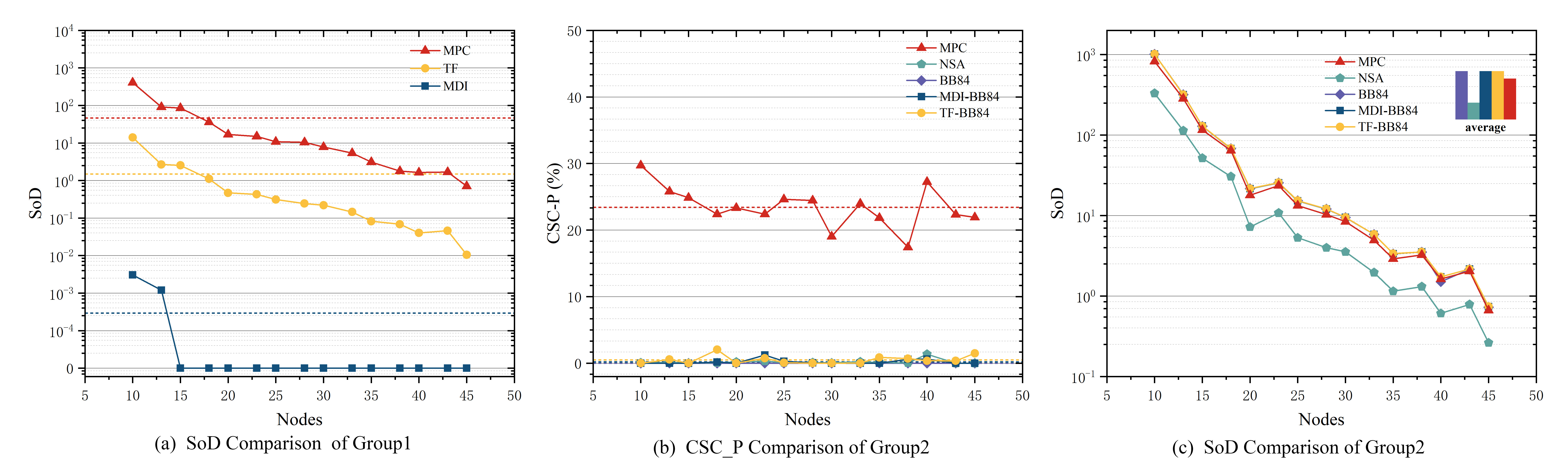}
	\caption{The results of experiment. Since all the flows in group 1 QKD networks are CSC-flows, the value of CSC\_P for all QKD networks in Group 1 is 100\%.}\label{fig4}
\end{figure*}

The comparison results of Group 1 in Figure~\ref{fig4}a show that the SoD of MPC QKD networks significantly outperforms that of TF QKD networks and MDI QKD networks across all networks scales. The SoD of MDI QKD networks is the lowest due to the limited key rate of the MDI protocol. Furthermore, the SoD of MDI QKD networks gradually decreases towards 0 as the network scale increases. Since the key rate of TF protocol is higher than that of the MDI protocol, TF QKD networks can still provide communication services when the network scale is slightly larger. However, the simulated results show that when the number of network nodes exceeds 18, the SoD falls below 1, indicating that it cannot satisfy the communication demands of all communication pairs. In contrast, the MPC QKD networks exhibit significantly better performance compared to the MDI and TF QKD networks. Even as the network scale increases, the MPC QKD networks maintain a SoD value greater than 1 (which means that it can satisfy the communication demands of all communication pairs) until the number of network nodes exceeds 43. The dotted line in the figure represents the mean value, clearly showing that the SoD of the MPC QKD networks is obviously higher than that of TF and MDI QKD networks, indicating that the proposed MPC QKD scheme has better communication capability.

The CSC\_P comparison results of Group2 in Figure~\ref{fig4}b show that the CSC\_P of the MPC QKD networks is significantly higher than that of other QKD networks, implying a weaker dependence on strongly trusted relays. Furthermore, the results also show the drawbacks of the BB84-MDI and BB84-TF QKD networks. These hybrid protocol QKD networks predominantly rely on the BB84 protocol for traffic transmission, with a very low proportion of traffic passing through untrusted relays. This indicates that these hybrid protocol QKD networks do not effectively reduce the dependence on strongly trusted relays in the QKD networks. In contrast, the proposed MPC QKD networks successfully reduce the dependence on strongly trusted relays. The simulated results in Figure~\ref{fig4}c show that the SoD of the MPC QKD networks is comparable to that of the BB84, BB84-MDI, and BB84-TF QKD networks. This indicates that the introduction of weak trusted nodes does not lead to a significant degradation of SoD. In contrast, the SoD of the NSA QKD networks is lower compared to other QKD networks, illustrating that the communication capability of the network can be improved through the collaboration of multiple protocols.

\section{Conclusion}
This paper proposes a multi-protocol collaborative networking scheme, providing a novel approach for constructing QKD networks. The first contribution of this scheme is the introduction of the MPC networking cell, which enhances the practicality of QKD networks. Additionally, to further enhance the communication capability, the problem of topology optimization is addressed. Simulated results show that the proposed scheme improves the communication capability by 37 times compared to the measurement device independent scheme and reduces the dependence on strongly trusted relays by 23\% compared to the measurement device dependent scheme. These results demonstrate the significant improvement in communication capability and the reduced dependence on strongly trusted relays achieved by the proposed multi-protocol collaborative networking scheme. This scheme holds great significance for advancing the practicalization of QKD networks.

\section*{Acknowledgments}
This work is supported by the National Natural Science Foundation of China (grant number: 62071151)


\bibliographystyle{IEEEtran}
\bibliography{bibfile}


 




\vfill

\end{document}